\documentstyle[aps,prl,floats,epsf,twocolumn]{revtex}

\begin{document}
\draft
\preprint{HEP/123-qed}
\newcommand{\dm}       {\Delta m^2}
\newcommand{\sinq}      {sin^2 2\theta}

\wideabs{
\title{Search for Neutrino Oscillations at 
the Palo Verde Nuclear Reactors
}

\author{F.~Boehm$^3$, J.~Busenitz$^1$, B.~Cook$^3$, G.~Gratta$^4$,
        H.~Henrikson$^3$, J.~Kornis$^1$, D.~Lawrence$^2$,
        K.B.~Lee$^3$, K.~McKinny$^1$, L.~Miller$^4$, 
        V.~Novikov$^3$, A.~Piepke$^3$, B.~Ritchie$^2$, D.~Tracy$^4$, 
        P.~Vogel$^3$, Y-F.~Wang$^4$, J.~Wolf$^1$}
\address{$^1$ Department of Physics and Astronomy, University of Alabama, Tuscaloosa AL 35487 \\
         $^2$ Department of Physics and Astronomy, Arizona State University, Tempe, AZ 85287 \\
         $^3$ Division of Physics, Mathematics and Astronomy, Caltech, Pasadena CA 91125 \\
         $^4$ Physics Department, Stanford University, Stanford CA 94305}

\date{\today}
\maketitle
\begin{abstract}

We report on the initial results from a measurement of the anti-neutrino
flux and spectrum at a distance of about 800~m from the three reactors
of the Palo Verde Nuclear Generating Station using a segmented gadolinium-loaded
scintillation detector.    We find that the anti-neutrino flux agrees with that 
predicted in the absence of oscillations 
excluding at 90\%~CL 
$\rm\bar\nu_e - \bar\nu_x$ oscillations with $\Delta m^2 > 1.12\times 10^{-3}$~eV$^2$
for maximal mixing and $\sin^2{2\theta} > 0.21$ for large $\Delta m^2$.

\end{abstract}


}

\narrowtext


Nuclear reactors have been used as intense sources
of $\bar\nu_{\rm e}$ in experiments searching for neutrino 
oscillations~\cite{hist}. 
These experiments usually detect $\rm \bar\nu_e$ by the 
process $\rm \bar\nu_e + p \rightarrow n + e^+$, where 
the cross-section-weighted energy spectrum of $\bar\nu_{\rm e}$, 
peaking at about 4 MeV, can be deduced from the measured $\rm e^+$ spectrum.
Any $\rm \bar\nu_e$ flux deficit or distortions of the $\rm \bar\nu_e$ 
energy spectrum would indicate oscillations.   
The low energy of reactor $\bar\nu_{\rm e}$ 
allows these experiments to reach very small mass parameters, albeit with 
modest mixing-angle sensitivity.
Past experiments~\cite{za} with detectors at 50-100~m from a reactor 
have explored the mass-parameter range down to $\rm 10^{-2}~eV^{2}$.
The work described here, and a similar experiment elsewhere~\cite{chooz},
are the first long baseline ($\sim$1~km) searches, designed to explore the parameter range
down to $\rm 10^{-3}~eV^{2}$ as suggested by the early Kamiokande atmospheric
neutrino anomaly~\cite{kamiokande}. Although later results from 
Super-Kamiokande~\cite{SK} (appeared while this work was in progress)
seem to disfavor the $\nu_{\mu} - \nu_{\rm e}$ channel, a direct experimental
exploration amply motivated this work.

\begin{figure}[tbh!!!!!]
\centerline{\epsfxsize=2.7in \epsfbox{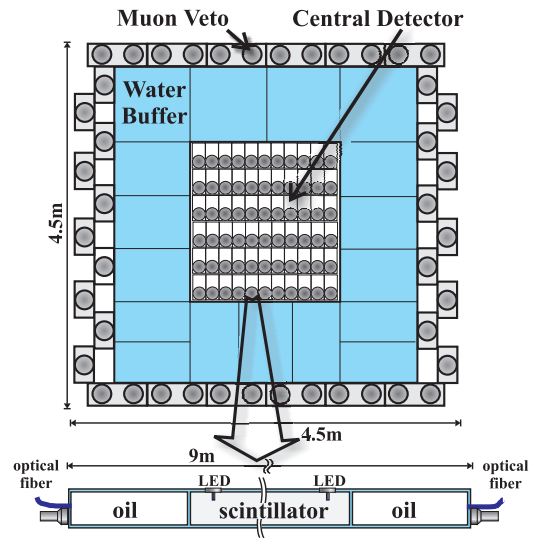}}
\caption{Schematic view of the Palo Verde neutrino detector.}
\label{fig:detector}
\end{figure}

The Palo Verde neutrino oscillation experiment is located at the Palo
Verde Nuclear Generating Station near Phoenix, Arizona. The total thermal 
power from three identical pressurized water reactors is 11.6~GW.      
Two of the reactors are 890~m from the detector, while the third is at 750~m. 
Our detector is placed in a shallow underground site (32~mwe overburden), 
thus eliminating the hadronic component of cosmic radiation and reducing 
the muon flux by a factor of $\sim 5$.
The fiducial mass, segmented to reject the remaining background, consists of 
11.3 tons of 0.1\% Gd-loaded liquid scintillator contained in a $6\times 11$ 
array of 9 m-long acrylic cells, as shown in Fig.~\ref{fig:detector}.  
Each cell is viewed by two 5-inch photomultiplier tubes, one at each end.
A $\rm \bar\nu_e$ is identified by space- and time-correlated 
$\rm e^+$ and n signals.
Positrons deposit their energies in the scintillator and annihilate,
yielding two 511 keV $\gamma$'s, giving a triple coincidence.
Neutrons thermalize and  are captured in Gd, giving a $\gamma$-ray shower 
of 8 MeV total energy. 

The Gd loading of the scintillator has two advantages: it reduces the neutron capture
time from 170~$\mu$s (on protons) to 30~$\mu$s and provides a high energy 
$\gamma$ shower to tag the neutron capture, resulting in a substantial 
background reduction.  Both the positron and the neutron are triggered by a 
triple coincidence requiring at least one cell above a ``high'' threshold set 
at about 600~keV (positron ionization or neutron shower core),
and two cells above a ``low'' threshold set at about 40~keV 
(Compton scattering from annihilation photons or neutron 
shower tails). The triple coincidences are required to be within a 
$3\times 5$ matrix anywhere in the detector. 

The central detector is surrounded by a 1~m water 
shield to moderate background neutrons produced by muons outside the detector 
and to absorb $\gamma$'s from the laboratory walls.
Outside the water tanks  are 32 large liquid scintillator counters
and two end-caps to veto cosmic muons. The  rate of cosmic muons is 
approximately 2~kHz. The pattern of muons traveling through 
veto chambers and their timing relative to the central detector 
hits are recorded for subsequent off-line analysis.
The central detector is equipped with a system of tubes that allows the insertion
of calibration sources in the small spaces between cells.   
In addition, a set of blue LEDs and optical fibers can produce flashes of light 
inside each of the cells.
In order to reduce natural radioactivity, all building materials for the 
detector are carefully selected, including the aggregate (marble) used in the
concrete of the underground  laboratory.

The detector geometry, materials and electromagnetic interactions are simulated 
using the package GEANT.
Hadronic interactions are described by FLUKA
and the low energy neutron transport is simulated by 
GCALOR.
The inclusive gamma spectrum from neutron 
capture on Gd is specially modeled according to measurements~\cite{gd}.
Light quenching effects are also included~\cite{quench}.


Since the ultimate sensitivity of the experiment relies on a disappearance
measurement, precise knowledge of the detector efficiency and of the
expected $\bar\nu_{\rm e}$ flux from the reactors is essential.

The efficiency calibration is based upon a primary measurement 
performed a few times per year with a calibrated $^{22}$Na $\rm e^+$ source and 
an Am-Be neutron source.
The $^{22}$Na source is placed in the calibration pipes and mimics 
the effects of the positron from the $\rm \bar\nu_e$ interaction by providing 
annihilation radiation and a 1.275~MeV photon which simulates the 
e$^+$ ionization in the scintillator.   The source is 
placed at 18 positions in the detector deemed to be representative of 
different conditions. 
The neutron detection efficiency is measured by scanning the detector with the
Am-Be source where the 4.4~MeV $\gamma$ associated with the neutron emission is tagged
with a miniaturized NaI(Tl) counter.  

Other calibrations, used to measure the detector energy response,  are 
performed using the Compton edges from $^{137}$Cs, $^{65}$Zn and $^{228}$Th
sources.     The same Th source is also used more frequently to track the
scintillator transparency.
Weekly runs of the fiber-optic and LED flasher systems are used, respectively,
to monitor the gain and linearity of photomultipliers and the 
timing/position relationship along the cells.    

Since the  energy deposition of the 511~keV $\gamma$'s in one cell 
has a sharp falling spectrum (Compton scattering) it is vital to have 
the lowest possible ``low'' thresholds in the trigger and to understand 
the behavior of  such thresholds with great accuracy.
This second task is complicated by the fact that the trigger uses 
voltage amplitudes, while only charge from integrating ADCs is available off-line.
For this reason our detector simulation includes a  detailed  
description of the signal development in time.
This code correctly describes the shape of pulses taking into 
account scintillator light yield, attenuation length and 
de-excitation time; photomultiplier rise- and fall-time and gain; and event
position along a cell.
The simulation of the detector response to the $^{22}$Na source correctly
describes the 40~keV (600~keV) threshold position to within 1.4~keV (2.6~keV),
resulting in an uncertainty on the positron (neutron) efficiency of 4\% (3\%). 

The $\bar\nu_{\rm e}$ flux and spectrum from a fission reactor and
the $\rm \bar\nu_e + p \rightarrow n + e^+$ cross section are
well known~\cite{hist,za,petr_john} and are calculated by tracking the 
$^{235}$U, $^{238}$U, $^{239}$Pu, and $^{241}$Pu fission rates 
in the three plant reactors, taking into account 
both power level and fuel age. 
%
%
The uncertainty in the $\rm\bar\nu_e$ reaction rate is less than 3\%.


The data presented here was collected in periods of 67.3 days
in 1998 and 134.4 days in 1999.   During the 98 (99) data taking one 
of the far (near) reactors was off for 31.3 (23.4) days.
While a detailed description of the data analysis will be reported 
elsewhere\cite{pv_prd}, here we will outline the principles of the
analysis and the results.

Neutrino candidates were selected by requiring an appropriate pattern
of energy to be present in the detector for the positron- and the neutron-like parts of the
events.    In addition the two sub-events are required to occur closer than about 1~m
from each other.   

At our depth the background to $\rm \bar\nu_e$ events consists of 
two types of events:  uncorrelated hits from cosmic-rays and natural 
radioactivity and correlated ones from cosmic-muon-induced neutrons. 
The first type can be measured by studying the 
time difference between positron-like and neutron-like parts of 
an event. 
%
%
By requiring that the time lapse between the two sub-events $t_{\rm en}$
be $5~\mu{\rm s} < t_{\rm en} < 200~\mu{\rm s}$, the uncorrelated background is 
reduced to $1.9\pm 0.1$~events~d$^{-1}$ ($3.7\pm 0.2$~events~d$^{-1}$) for 
1998 (1999), as measured from a fit to the capture exponential and a constant.

The distribution of time intervals between a cosmic-ray $\mu$ crossing the 
detector and a $\bar\nu_{\rm e}$-like event carries information on the 
correlated background.   From an exponential fit 
we infer that the majority of correlated background is produced by pairs
of neutrons, where the capture of the first neutron in each pair mimics the 
positron signature.   
The requirement that no cosmic-ray hits be present in a window of $150~\mu$s
preceding the $\bar\nu_{\rm e}$ candidate completes the event selection.
While the detector efficiency, $\eta$, is dependent upon the neutrino energy, the 
efficiency integrated over the neutrino spectrum
produced by the reactor (i.e. in the case of no oscillation) is $\simeq 8\%$
($\simeq 11\%$) for 1998 (1999) (the higher efficiency in 99 is due to improvements 
in the data acquisition dead-time and trigger efficiency).
%
%

\begin{figure}[hb!!!!]
\centerline{
\epsfxsize=3.25in \epsfbox{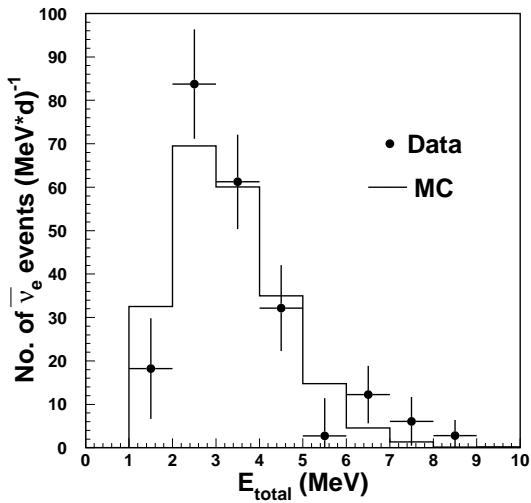}}
\caption{Spectrum of energy deposited for neutrino events
after ``on''-``off'' background subtraction.
The sum of the 1998 and 1999 distributions is compared with the Monte Carlo 
simulation assuming no oscillations. Uncertainties are statistical only.
The $\chi^2/{\rm n.d.f.}$ between simulation and data is $8.5/8$, supporting 
the no-oscillation scenario.}
\label{fig:e_spectrum}
\end{figure}

The resulting rates $N$ of $\bar\nu_{\rm e}$ candidates per day in different
periods are given in Tab.~\ref{tab:results}.
Along with the $\bar\nu_{\rm e}$ events this final data set contains the random 
background mentioned above and a substantial amount of correlated background.
Two independent techniques were used to estimate and subtract the 
background.
The most straightforward method (``Method 1'') relies on the changes of 
the $\bar\nu_{\rm e}$
signal when different reactors are turned off.   For simplicity we first 
directly subtract the efficiency-corrected rates $r=N/\eta$ for 
``off'' periods from the ``on'' periods.   
For 1998 (1999) we compare the efficiency-corrected rates of observed 
interactions $95.0\pm 18.6$~d$^{-1}$ ($76.8\pm 13.6$~d$^{-1}$) with
the prediction of 63~d$^{-1}$ (88~d$^{-1}$).    While the uncertainties quoted
here are statistical only, we  find  good agreement with the 
hypothesis of no oscillations.    We can then proceed to add the 1998 and 1999
data sets (which have a different combination of baselines) and obtain a 
neutrino energy spectrum for the difference ``on''-``off''.   This spectrum
well matches the model for no oscillations as shown in Fig.~\ref{fig:e_spectrum}.       

\begin{figure}[htb!!!!!!]
\centerline{
\epsfxsize=3.35in \epsfbox{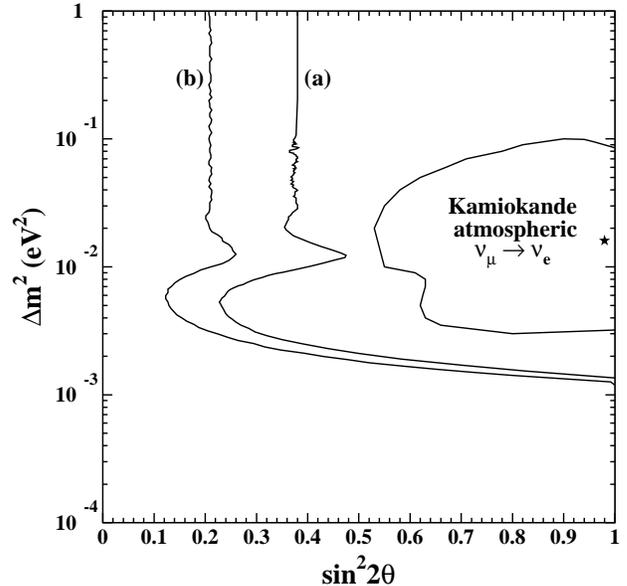}}
\caption{Limits on mass difference and mixing angle from the present work (90\%~CL).
Curves (a) and (b) are based on Method 1 and 2 for background subtraction, respectively,
as described in the text. The method~[10] is used to calculate the
exclusion contours. The Kamiokande $\rm \nu_{\mu} - \nu_e$ atmospheric neutrino
result is also shown.}
\label{fig:excl_plot}
\end{figure}

In order to test quantitatively the oscillation hypothesis in the $\dm - \sinq$ 
plane we  perform a $\chi ^2$-analysis comparing the expected   
to the measured rates. Information on the reactor neutrino flux, in the form of 
burn-up dependent fission rates, is computed for every detector run
(on average $\sim$12~h of data).   This analysis effectively unfolds the 
background from the data using the $\rm\bar\nu_e$-flux variation due to reactor
power changes and fuel burn-up. At each point of the $\dm - \sinq$ grid the 
energy spectrum and the detector efficiency are calculated and used to predict 
the expected rates~$\rho_i$, where $i$ denotes the different 
detector runs.    We then construct the function:
\begin{equation}
\chi^2=\sum_{i} \frac{((\alpha \rho_i + b)-N_i)^2}{\sigma_i^2}+\frac{(\alpha-1)^2}{\sigma_{\rm syst}^2}.
\end{equation}
The $\chi ^2$ is 
minimized with respect to the background contribution $b$ to the candidate neutrino rate $N_i$ 
and the parameter $\alpha$ that accounts for possible global normalization effects.
While statistical errors $\sigma_i$ have to be individually applied to each run, the 
systematic error $\sigma_{\rm syst}$ is treated as a global parameter in the $\chi ^2$.
The contribution of different sources of systematics to $\sigma_{\rm syst}$ is given
in the first column of Tab.~\ref{tab:systematics}.    The effect of the $\rm\bar\nu_e$ 
selection cuts is estimated by a random sampling of the unity hyper-volume defined by 
conservative ranges for each of the individual cuts. This technique properly takes 
into account the possible correlations between cuts. 
The 90\%~CL acceptance region is defined in accordance with the procedure 
in~\cite{F_C}
by $\Delta \chi^2 > \chi^2(\dm,\sinq) - \chi^2_{\rm best}$ where $\chi^2(\dm,\sinq)$ 
describes the fit quality at the current grid point and $\chi^2_{\rm best}$ that 
of the global best fit determined in the physically allowed parameter space. 
This procedure clearly prefers a no-oscillation scenario as illustrated in 
Fig.~\ref{fig:excl_plot}~(curve a).    The fit also provides an estimate of the background
$b = 19.5 \pm 1.3$~d$^{-1}$ ($b = 26.3 \pm 1.7$~d$^{-1}$) for 1998 (1999),
and of $\alpha = 1.02 \pm 0.08$ with a $\chi^2/{\rm n.d.f.} = 318/327$.
%
\begin{table}
\begin{tabular}{|l|c|c|c|c|}
Period                         & 98 ``on'' & 98 ``off''$^{\ast}$ & 99 ``on'' & 99 ``off''$^{\dagger}$  \\
\hline
Duration (d)              & 36.0      & 31.3                &  111.0    &   23.4               \\
$\eta$                    & 0.0746    & 0.0772              &  0.112    &   0.111              \\
$N$ (d$^{-1}$)            & $38.2\pm 1.0$ & $32.2\pm 1.0$ & $52.9\pm 0.7$    & $43.9\pm 1.4$  \\
\hline
$S_{\nu}$ (d$^{-1}$)      & $16.5\pm 1.4$ & $13.4\pm 1.4$ & $25.2\pm 0.9$    & $15.1\pm 1.9$  \\
$B$ (d$^{-1}$)            & $21.7\pm 1.0$ & $18.8\pm 1.0$ & $27.7\pm 0.6$    & $28.8\pm 1.3$  \\
$R_{\rm Obs}$ (d$^{-1}$)  & $221\pm 18$   & $174\pm 17$   & $225\pm 8$       & $136\pm 17$    \\
$R_{\rm Calc}$ (d$^{-1}$) & $218$         & $155$         & $218$            & $130$          \\
\end{tabular}
\caption{Summary of results from the Palo Verde experiment.  The values in the second part of the
table are derived from Method 2.  $B=B_{\rm unc} + B_{\rm nn} + B_{\rm np}$. 
$R_{\rm Obs}$ and $R_{\rm Calc}$ are the observed and calculated $\bar\nu_{\rm e}$ rates corrected 
by the efficiencies $\eta$ for the case of no-oscillations.  $^{\ast}$ Reactor at 890~m distance off. 
$^{\dagger}$ Reactor at 750~m distance off. Statistical uncertainties only.}
\label{tab:results}
\end{table}

In an independent analysis (``Method 2'') that will be described in detail 
elsewhere~\cite{swap} we make use of the intrinsic symmetry of the dominant 
two-neutron background to cancel most of the background directly from data 
and compute the remaining components from Monte Carlo simulations.   
This technique makes the best possible use of the statistical power of all 
data collected.
The rate of candidate events after all cuts can be written as 
$N = B_{\rm unc} + B_{\rm nn} + B_{\rm pn} + S_{\nu}$ where 
the contribution of the uncorrelated $B_{\rm unc}$, two-neutron $B_{\rm nn}$ and 
other correlated backgrounds $B_{\rm pn}$ are explicitly represented, along 
with the $\bar\nu_{\rm e}$ signal $S_{\nu}$.   
The dominant background $B_{\rm nn}$ (along with $B_{\rm unc}$)
is symmetric under exchange of sub-events, so that an event selection with the 
requirements for the prompt and delayed event parts swapped will result in a rate 
$N^{\prime} = B_{\rm unc} + B_{\rm nn} + \epsilon_1 B_{\rm pn} + \epsilon_2 S_{\nu}$
where $\epsilon_1$ and $\epsilon_2$ account for the different efficiency for 
selecting asymmetric events after the swap.
\begin{table}
\begin{tabular}{|l|c|c|}
Systematic                        & Method 1 (\%) & Method 2 (\%) \\
\hline
e$^+$ efficiency                  &       4       &       4       \\
n  efficiency                     &       3       &       3       \\
$\bar\nu_{\rm e}$ flux prediction &       3       &       3       \\
$\bar\nu_{\rm e}$ selection cuts  &       8       &       4       \\
$B_{\rm pn}$ estimate             &      ---      &       4       \\
\hline
Total                             &      10       &       8       \\
\end{tabular}
\caption{Origin and magnitude of systematic errors.
Using Method 2 for background subtraction reduces the systematic uncertainty
from the event selection cuts but introduces a new uncertainty due to
the accuracy of the Monte Carlo used for the estimate of $B_{\rm pn}$.}
\label{tab:systematics}
\end{table}
We then calculate $N - N^{\prime} = (1-\epsilon_1)B_{\rm pn} + (1-\epsilon_2)S_{\nu}$
where the efficiency correction $\epsilon_2\simeq 0.2$ can be estimated from 
the $\rm\bar\nu_e$ Monte-Carlo simulation.
We find that the processes of $\mu$-spallation in the laboratory walls and capture
of the $\mu$'s that are not tagged by the veto counter ($4\pm 1$)\% contribute
to $(1-\epsilon_1) B_{\rm pn}$, while other backgrounds are negligible.
Using Monte-Carlo simulation, we obtain $(1-\epsilon_1)B_{\rm pn}=-0.9\pm 0.5$~d$^{-1}$
($-1.3\pm 0.6$~d$^{-1}$) for $\mu$-spallation in 1998 (1999); the same figures
for $\mu$-capture are $0.6\pm 0.3$~d$^{-1}$ ($0.9\pm 0.5$~d$^{-1}$) in 1998 (1999).
This represents only a small correction to $N-N^{\prime}$ since 
$\epsilon_1$ is close to 1.  While the Monte-Carlo model is accurate for the capture
process, in the case of spallation we simulate the broad range of spectral indexes for 
the n-recoil energy reported in literature~\cite{swap}.  The average between different 
predictions is then used for $B_{\rm pn}$ while the spread is used as an extra 
systematic error in the second column of Tab.~\ref{tab:systematics}.
Since no $\bar\nu_{\rm e}$ signal is present above 10~MeV, the observed integrated
rate above such energy is used as a normalization of the Monte Carlo. 
The results  are shown  
in the second part of Tab.~\ref{tab:results} for different running periods.     
Clearly Method 2 is also in agreement with the no-oscillation hypothesis.
The excluded region, calculated by comparing the expected and observed $\rm\bar\nu_e$
rates taking into account the effect of the oscillation parameters on $\eta$ and $\epsilon_2$,
is given in Fig.~\ref{fig:excl_plot}~(curve b).


In conclusion, the data from the first period of running from the Palo Verde detector 
shows no evidence for $\rm \bar\nu_e - \bar\nu_x$ oscillations.
This result, together with the data already reported by Super-Kamiokande~\cite{SK} 
and Chooz~\cite{chooz}, excludes the channel $\rm \nu_{\mu} - \nu_e$ as being responsible 
for the atmospheric neutrino anomaly reported by Kamiokande~\cite{kamiokande}.
Data-taking at Palo Verde is scheduled to continue until the summer of 2000.


We would like to thank the Arizona Public Service Company for
the generous hospitality at the Palo Verde plant.
The important contributions of M.~Chen, R.~Hertenberger, K.~Lou and 
N.~Mascarenhas in the early stages of this project are gratefully 
acknowledged.
We are indebted to J.F.~Beacom, B.~Barish, R.~Canny, A.~Godber, 
J.~Hanson, D.~Michael, C.~Peck, C.~Roat, N.~Tolich and A.~Vital 
for their help.
We also acknowledge the generous financial help from
the University of Alabama, Arizona State University, Caltech and 
Stanford University.     Finally our gratitude goes to CERN, DESY, FNAL,
LANL, LLNL, SLAC and TJNAF who at different times provided us with 
parts and equipment for the experiment.
This project was supported in part by the US DoE. 
One of us (J.K.) also received support from the Hungarian OTKA fund.

\end{document}